\documentclass[conference]{IEEEtran}
\IEEEoverridecommandlockouts
\usepackage{graphicx}
\usepackage{placeins}
\usepackage{hyperref}
\usepackage{amsmath}
\usepackage{pifont}
\usepackage{float}
\usepackage{llncsconf}
\usepackage{array}
\usepackage{multirow}

\usepackage{multicol}
\usepackage[per-mode=symbol]{siunitx}
\usepackage{comment}
\usepackage{siunitx}
\usepackage{algorithm}
\usepackage{amsmath, amsthm}
\usepackage{xurl}
\usepackage{algpseudocode}

\usepackage{subcaption}
\usepackage[inline,shortlabels]{enumitem}
\usepackage{bbding}
\usepackage{siunitx}
\usepackage{url}

\usepackage{tikz}
\usepackage{cite}
\usepackage{amsmath,amssymb,amsfonts}
\usepackage{textcomp}
\usepackage{xcolor}
\usepackage{booktabs}
\usepackage{fontawesome}
\algblock{Input}{EndInput}
\algnotext{EndInput}
\algblock{Output}{EndOutput}
\algnotext{EndOutput}
\newcommand{\Desc}[2]{\State \makebox[2em][l]{#1}#2}

\def\BibTeX{{\rm B\kern-.05em{\sc i\kern-.025em b}\kern-.08em
    T\kern-.1667em\lower.7ex\hbox{E}\kern-.125emX}}

\algdef{SE}[VARIABLES]{Variables}{EndVariable}
   {\algorithmicvariables}
   {\algorithmicend\ \algorithmicvariables}
\algnewcommand{\algorithmicvariables}{\textbf{Global variables}}

\newcommand{\var}{\texttt}

\newcommand{\Parameter}[2]{\Statex $\triangleright$ \var{#1}: #2}

\begin{document}

\title{SCAREY: Location-Aware Service Lifecycle Management}

\makeatletter
\newcommand{\newlineauthors}{%
  \end{@IEEEauthorhalign}\hfill\mbox{}\par
  \mbox{}\hfill\begin{@IEEEauthorhalign}
}

\newcommand{\WRP}{\par\qquad\qquad\qquad\qquad\qquad\(\hookrightarrow\)\enspace}
\newcommand{\rnode}{\textsc{$Reference City$}}

\newcommand\copyrighttext{%
  \footnotesize \textcopyright \the\year{} IEEE. Personal use of this material is permitted. Permission from IEEE must be obtained for all other uses, including reprinting/republishing this material for advertising or promotional purposes, collecting new collected works for resale or redistribution to servers or lists, or reuse of any copyrighted component of this work in other works.}

\newcommand\copyrightnotice{%
\begin{tikzpicture}[remember picture,overlay]
\node[anchor=south,yshift=10pt] at (current page.south) {\fbox{\parbox{\dimexpr0.75\textwidth-\fboxsep-\fboxrule\relax}{\copyrighttext}}};
\end{tikzpicture}%
}


\makeatother

\author{\IEEEauthorblockN{Kurt Horvath}
\IEEEauthorblockA{\textit{Institute of Information Technology} \\
\textit{University of Klagenfurt, Austria}\\
0009-0008-7737-7013 \\
kurt.horvath@aau.at}
~\\
\and
\IEEEauthorblockN{Dragi Kimovski}
\IEEEauthorblockA{\textit{Institute of Information Technology} \\
\textit{University of Klagenfurt, Austria}\\
0000-0001-5933-3246 \\
dragi.kimovski@aau.at }
~\\
\and
\IEEEauthorblockN{Radu Prodan}
\IEEEauthorblockA{\textit{Department of Computer Science} \\
\textit{University of Innsbruck, Austria}\\
0000-0002-8247-5426 \\
radu.prodan@uibk.ac.at}

}

\maketitle
\copyrightnotice

\begin{abstract}
Scheduling services within the computing continuum is complex due to the dynamic interplay of the Edge, Fog, and Cloud resources, each offering distinct computational and networking advantages. This paper introduces SCAREY, a user location-aided service lifecycle management framework based on state machines. SCAREY addresses critical service discovery, provisioning, placement, and monitoring challenges by providing unified dynamic state machine-based lifecycle management, allowing instances to transition between discoverable and non-discoverable states based on demand. It incorporates a scalable service deployment algorithm to adjust the number of instances and employs network measurements to optimize service placement, ensuring minimal latency and enhancing sustainability. Real-world evaluations demonstrate a \qty{73}{\percent} improvement in service discovery and acquisition times, \qty{45}{\percent} cheaper operating costs and over \qty{57}{\percent} less power consumption and lower CO\textsubscript{2} emissions compared to existing related methods.

\begin{IEEEkeywords}
Computing Continuum, Life Cycle Management, Scheduling
\end{IEEEkeywords}
\end{abstract}

\section{Introduction}\label{sec:Introduction}

Scheduling of services in the computing continuum (CC) is a complex and evolving problem. Typically, application owners selectively deploy Edge service instances, while Fog and Cloud resources provide complementary but distinct coverage and backup. With demand fluctuating throughout the day, optimized scheduling configurations vary dynamically. Recent years have seen the development of various methods tailored to specific scenarios and constraints, aiming to establish what constitutes the ``optimal'' scheduling of Internet of Things (IoT) applications \cite{Mehran2024SchedulingSurvey,Taghinezhad-Niar2024SecurityEnvironments}. These constraints often include limitations on the latency \cite{Chiaro2024Latency-awareContinuum} or the maximum time allowed for application deployment.

Additionally, sustainability and economic considerations are becoming increasingly important. When demand falls below a certain threshold, one can turn off unused resources, such as Edge devices and Cloud instances, as long as another service instance remains available to serve users. This temporal and spatial volatility of service instances managed through lifecycle introduces new challenges beyond scheduling. Key domains, such as service discovery, provisioning, service placement, and monitoring, must adopt a unified approach to meet quality of service constraints effectively.

Each domain -- service discovery, provisioning, placement, and monitoring -- has developed numerous solutions that address various aspects of complete lifecycle management \cite{Saito2024AContinuum,Heidari2022ServiceReview}. Therefore, creating a unified method that provides a holistic solution for service operators, encompassing service owners, providers, and end consumers, ensures consistent fulfilment of all quality of service (QoS) metrics, regardless of fluctuations in demand or shifts in service instance location and availability. Integrating these domains under a cohesive management strategy simplifies operations and improves the reliability of services across the computing continuum.

This paper introduces SCAREY, a \textit{uSer loCation Aided seRvice lifecYcle} end-to-end solution for defining and managing the lifecycle of services deployed within a resource-constrained computing continuum using state machines for demand-driven service instances management. Service instances can be in different states depending on their visibility to the user. Increasing demand will make a service instance discoverable and turn it undiscoverable if demand declines, balancing users to other instances. Therefore, SCAREY provides four innovative design characteristics.

\paragraph{State machine-based lifecycle management} based on user demand that triggers transitions, guiding each service instance from stored to final state and instance removal.

\paragraph{Scalable service deployment} and dynamic service scaling algorithm adjusting the number of instances responding and meeting user demands.

\paragraph{Service placement optimization} with minimized latency and enhanced user satisfaction using network measurements to evaluate the performance, reliability, and suitability of Edge, Fog, and Cloud resources for hosting services.

\paragraph{Selective resource acquisition} reducing service idle time based on user demand and location, resource use and greenhouse gas emissions, particularly carbon dioxide (CO\textsubscript{2}).

The paper has eight sections. Section \ref{sec:relatedWork} overviews the existing works and their relevance to our methodology. Section \ref{sec:Model}, formally describes SCAREY methodology. Section \ref{sec:problem} discusses the problem based on the established model, followed by Section 
\ref{sec:framework} describing the architecture and the required components to operate. Section \ref{sec:kpi} defines the testbed and evaluation scenarios, and Section \ref{sec:Results} discusses the results. We conclude by summarizing the results in Section \ref{sec:Conclusion}.

\section{Related Work}\label{sec:relatedWork}
This section examines the related approaches for service life cycle management. Table \ref{tab:sota} summarizes eight methods that support life cycle management, including discovery, scheduling, and deployment \cite{Kimovski2021CloudCompute}.
Our categorization of the related methods considers multiple aspects, including latency awareness, resource utilization and optimization, and cost benefits. When assessing latency and system resource usage, we further distinguish between approaches that directly measure these aspects and estimate them through analytical models. 


\begin{table*}[t]
\centering

\resizebox{.9\textwidth}{!}{
\begin{tabular}{lc|cccc|ccc|c|}
\cline{3-9}
 & \multicolumn{1}{l|}{} & \multicolumn{4}{c|}{\textbf{Network Latency}} & \multicolumn{3}{c|}{\textbf{System Load}} & \multicolumn{1}{l}{} \\ \hline
\multicolumn{1}{|l|}{\textbf{Framework}} & \textbf{Constraints} & \multicolumn{1}{c|}{\textbf{Used}} & \multicolumn{1}{c|}{\textbf{Measured}} & \multicolumn{1}{c|}{\textbf{Calculated}} & \textbf{Evaluation Method} & \multicolumn{1}{c|}{\textbf{Used}} & \multicolumn{1}{c|}{\textbf{Measured}} & \textbf{Estimated} & \multicolumn{1}{l|}{\textbf{Disabling Nodes}} \\ \hline
\multicolumn{1}{|l|}{LAIS \cite{Chiaro2024Latency-awareContinuum}} & \begin{tabular}[c]{@{}c@{}}Kubernetes-based, \\ multi-cluster required\end{tabular} & \multicolumn{1}{c|}{\Checkmark} & \multicolumn{1}{c|}{\Checkmark} & \multicolumn{1}{c|}{\XSolid} & Measured live & \multicolumn{1}{c|}{\XSolid} & \multicolumn{1}{c|}{\Checkmark} & \XSolid & \XSolid \\ \hline
\multicolumn{1}{|l|}{ATS-FOA \cite{Memari2022AArchitecture}} & \begin{tabular}[c]{@{}c@{}}Multi-layer Fog-Cloud \\ architecture required\end{tabular} & \multicolumn{1}{c|}{\Checkmark} & \multicolumn{1}{c|}{\XSolid} & \multicolumn{1}{c|}{\Checkmark} & \begin{tabular}[c]{@{}c@{}}Caculated by decimating\\ system latency by \\ processing latency\end{tabular} & \multicolumn{1}{c|}{\Checkmark} & \multicolumn{1}{c|}{\XSolid} & \Checkmark & \XSolid \\ \hline
\multicolumn{1}{|l|}{JDATP \cite{Tang2022Latency-awareSystems}} & \begin{tabular}[c]{@{}c@{}}OpenStack, \\ OpenDaylight, \\ Kubernetes\end{tabular} & \multicolumn{1}{c|}{\Checkmark} & \multicolumn{1}{c|}{\XSolid} & \multicolumn{1}{c|}{\Checkmark} & Based on system latency & \multicolumn{1}{c|}{\Checkmark} & \multicolumn{1}{c|}{\Checkmark} & \Checkmark & \Checkmark \\ \hline
\multicolumn{1}{|l|}{COLAP \cite{Gupta2017COLAP:Environment}} & \begin{tabular}[c]{@{}c@{}}Multi-Cloud \\ NFV deployment, \\ no live measurement\end{tabular} & \multicolumn{1}{c|}{\Checkmark} & \multicolumn{1}{c|}{\XSolid} & \multicolumn{1}{c|}{\Checkmark} & Modelled via SVR & \multicolumn{1}{c|}{\Checkmark} & \multicolumn{1}{c|}{\XSolid} & \Checkmark & \Checkmark \\ \hline
\multicolumn{1}{|l|}{PIES \cite{Hudson2021QoS-AwareImplementations}} & Edge AI-focused & \multicolumn{1}{c|}{\Checkmark} & \multicolumn{1}{c|}{\XSolid} & \multicolumn{1}{c|}{\Checkmark} & \begin{tabular}[c]{@{}c@{}}Estimated from \\ bandwidth and \\ user count\end{tabular} & \multicolumn{1}{c|}{\Checkmark} & \multicolumn{1}{c|}{\XSolid} & \Checkmark & \XSolid \\ \hline
\multicolumn{1}{|l|}{LAKS \cite{Centofanti2023Latency-AwareEdge}} & \begin{tabular}[c]{@{}c@{}}Kubernetes-based, \\ edge-focused, \\ MQTT required\end{tabular} & \multicolumn{1}{c|}{\Checkmark} & \multicolumn{1}{c|}{\Checkmark} & \multicolumn{1}{c|}{\Checkmark} & Measured live & \multicolumn{1}{c|}{\Checkmark} & \multicolumn{1}{c|}{\Checkmark} & \XSolid & \multicolumn{1}{c|}{\XSolid} \\ \hline

\multicolumn{1}{|l|}{CAS \cite{Tao2021ContainerComputing}} & \begin{tabular}[c]{@{}c@{}}Serverless-focused,\\ Kubernetes required\end{tabular} & \multicolumn{1}{c|}{\Checkmark} & \multicolumn{1}{c|}{\Checkmark} & \multicolumn{1}{c|}{\XSolid} & \begin{tabular}[c]{@{}c@{}}Indirectly estimated \\ via cold start \\ reduction\end{tabular} & \multicolumn{1}{c|}{\Checkmark} & \multicolumn{1}{c|}{\Checkmark} & \Checkmark & \multicolumn{1}{c|}{\XSolid} \\ \hline

\multicolumn{1}{|l|}{SCAREY} & \begin{tabular}[c]{@{}c@{}} Requires an external service discovery \\ and  latency measurements\\ performed in advance\end{tabular} & \multicolumn{1}{c|}{\Checkmark} & \multicolumn{1}{c|}{\Checkmark} & \multicolumn{1}{c|}{\XSolid} & Measured upfront & \multicolumn{1}{c|}{\Checkmark} & \multicolumn{1}{c|}{\XSolid} & \Checkmark & \Checkmark \\ \hline
\end{tabular}

}
\caption{Overview State of the Art}\label{tab:sota}
\vspace{-6mm}
\end{table*}

\paragraph{Latency-Aware Intent Scheduler (LAIS) \cite{Chiaro2024Latency-awareContinuum}}\label{par:lais} is a custom Kubernetes scheduler that dynamically places pods in multi-cluster environments based on real-time latency measurements. It monitors, records, and updates the network latency using \textit{Latency Meter}\cite{Centofanti2023Latency-AwareEdge} upon incoming user requests. LAIS prioritizes user-perceived quality of experience by adjusting pod placement to meet specified latency constraints. It operates in a closed-loop system, where the \textit{Descheduler} reallocates pods upon latency violations. Since LAIS gathers latency data live rather than upfront, it quickly adapts to changing network conditions and user mobility, ensuring optimized workload distribution.

\paragraph{Adaptive Tabu Search with Fruit Fly Optimization Algorithm (ATS-FOA) \cite{Memari2022AArchitecture}}\label{par:ats-foa}
presents a latency-aware scheduling heuristic for Fog environments, optimizing task allocation in smart home energy management \cite{Amadeo2019OnServices}. It estimates processing latency by measuring response and execution times, calculates network delays, continuously monitors task execution, and dynamically reallocates workloads to improve latency. The approximate nearest neighbor \cite{Arya1998AnDimensions} algorithm quickly identifies 'optimal' nodes, while FOA \cite{Wang2015AnProblems} refines scheduling decisions by avoiding congested or inefficient nodes.

\paragraph{Joint Data Access and Task Processing Algorithm (JDATP)\cite{Tang2022Latency-awareSystems}}\label{par:jdatp} categorizes latency into short-term (data access) and long-term (task processing) to optimize task execution in Edge and Cloud computing environments. JDATP considers transmission delay, disk I/O latency, and data retrieval time for short-term latency and selects servers that can quickly access required data and reduce transmission overhead. JDATP monitors network bandwidth and server processing times, reallocating tasks to avoid congestion. The algorithm calculates real-time network delays using queuing models, updating task scheduling decisions based on current network conditions, bandwidth, and traffic load.

\paragraph{Cost Optimised Latency Aware Placement (COLAP) \cite{Gupta2017COLAP:Environment}}\label{par:colab} actively considers network latency in service function chain placement across multiple Clouds. It integrates a queuing-theoretic model \cite{Shon2007ADetection} to estimate end-to-end latency using support vector regression \cite{Awad2015SupportRegression} to predict latency and minimizes active resources by strategically placing service chains as units, reducing inter-cloud communication overhead \cite{Huin2018OptimalProvisioning}. COLAP removes unnecessary nodes from the active pool in case of underutilized resources, preventing inefficient resource usage and lowering operational costs.


\paragraph{Placement of Intelligent Edge Services (PIES) \cite{Hudson2021QoS-AwareImplementations}}\label{par:pies} optimizes deep learning model placement across Edge and Clouds while considering QoS constraints, such as inference latency, model accuracy, and resource availability. PIES models transmission latency as a function of communication cost, Edge-Cloud bandwidth, and the number of user requests. It calculates network delay dynamically by distributing available bandwidth evenly among users. PIES places services based on demand to minimize resource consumption, ensuring it only deploys essential models. The algorithm avoids redundant deployments by placing only one instance per service where possible, preventing unnecessary resource use.

 \paragraph{Latency-Aware Kubernetes Scheduler (LAKS) \cite{Centofanti2023Latency-AwareEdge}}  \label{par:laks}
  optimizes microservice placement by dynamically measuring end-to-end application-layer latency. It deploys sentinel replicas across nodes, collects real-time round-trip time (RTT) from client interactions, and removes high-latency instances to optimize placement.
 LAKS gathers latency data using a meter application \cite{Centofanti2023Latency-AwareEdge}, which runs on both the client and server sides. It monitors RTT, publishes latency values via MQTT messages, and stores them in a model, tracking latency per pod and timestamps. In addition, LAKS considers CPU and memory availability to balance workloads when multiple nodes exhibit similar latency.

\paragraph{Container Lifecycle-Aware Scheduling (CAS) \cite{Tao2021ContainerComputing}}  \label{par:cas}
optimizes serverless computing by reducing cold start latency and improving resource allocation. The framework manages the entire container lifecycle, tracking state transitions from creation to eviction. CAS prioritizes reusing paused containers to minimize unnecessary container launches, reducing startup delay and network overhead. 
The framework dynamically schedules requests by selecting workers that already host suitable containers, avoiding the need to start new ones. CAS prevents premature eviction by preferring workers with available resources, reducing disruptions, and maintaining service continuity. By making proactive scheduling decisions, CAS minimizes resource waste \cite{Patros2021TowardComputing} and avoids unnecessary workload migrations contributing to network delays.


\paragraph{Ambition}\label{par:amb} 

SCAREY identifies improved service placements with minimal resource usage while ensuring the service meets its latency constraints. It dynamically adapts to demand, scaling down to a minimal resource footprint during low usage and scaling up seamlessly with increased demand. This flexibility enables modelling the complete lifecycle of service instances, offering a robust, large-scale solution for service discovery, provisioning, and deployment.  

\section{Lifecycle Model}\label{sec:Model}

We model the \emph{service lifecycle} as a finite state machine (FSM) \cite{Gill1962IntroductionMachines} $M = (Q, \Sigma, S_\mathtt{STO}, F)$, consisting of an \emph{initial service state} $S_\mathtt{STO}$ and
three finite sets:

\paragraph{Service states} $Q$ described in Section~\ref{sub:statemachine};
\paragraph{Service state transitions} $\Sigma$, where a \emph{transition} $T \in \Sigma$ is a function $T: Q \times \{ A \cup K \} \to Q$ that controls the instance deployment and visibility upon \emph{service maintenance events} $A$, or  \emph{demand constraints} $K$, defined in detail in Section \ref{sub:demandbasedtrans}.
\paragraph{Initial state $S_\mathtt{STO}$} described in Section~\ref{sub:statemachine};
\paragraph{Final service states} $F \subset Q$.

\begin{figure}[t]
    \centering
    \includegraphics[width=.5\columnwidth]{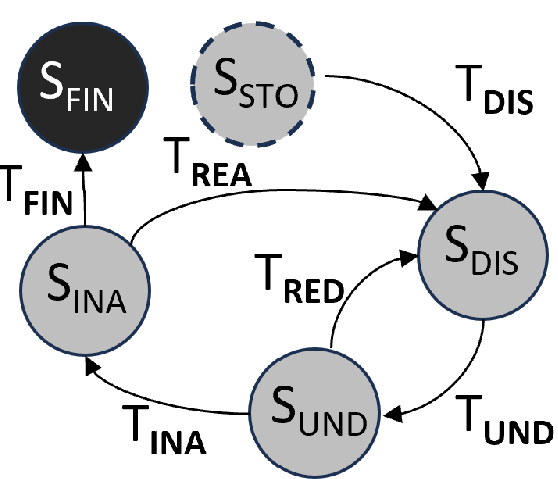} %
    \caption{Service state machine overview.}%
    \label{fig:overviewServicestate-machine}%
    \vspace{-6mm}
\end{figure}

\subsection{States}\label{sub:statemachine}
A service can be in five states: \mbox{$Q = \left\{ S_\mathtt{STO}, S_\mathtt{DIS}, S_\mathtt{UND}, S_\mathtt{INA}, S_\mathtt{FIN} \right\}$}.

\paragraph{Stored $S_\mathtt{STO}$} is the initial state indicating the complete configuration of the service in the form of an image, including all necessary parameters, dependencies, and resources. 

\paragraph{Discoverable $S_\mathtt{DIS}$} shows a running service from the stored image, accessible by users since the demand suffices.

\paragraph{Undiscoverable $S_\mathtt{UND}$} shows a low user demand and an unaccessible instance to the user.

\paragraph{Inactive $S_\mathtt{INA}$} serves as a temporary halt in service availability, reflecting a period dictated by temporal constraints such as maintenance, updates, or low-demand phases. 

\paragraph{Final $S_\mathtt{FIN}$} represents the end state, preventing any transitions to succeeding states, employed by the service operator for service decommissioning.

\subsection{Transition Types}\label{sub:transition}
A transition $T \in \Sigma$ can be of two kinds:

\subsubsection{Service maintenance transitions} $T(Q,A)$ control the service availability in two situations: $A=\left\{\alpha_\mathtt{INA}, \alpha_\mathtt{REA}\right\}$:

\paragraph{Inactive} $\alpha_\mathtt{INA}$\label{par:aIna} is an event triggered by a manual action from the service operator, which may involve making a service instance inaccessible, decommissioned, initiating maintenance work, or manually adjusting the service schedule.
\paragraph{Reactivate} $\alpha_\mathtt{REA}$\label{par:Rea} ensures users access to the service instance by reverting the previous changes to a discoverable state: $T\left(S,\alpha_\mathtt{REA}\right)=S_\mathtt{DIS}$. 

\subsubsection{Demand constraint transitions}\label{sub:demandbasedtrans} 
$T(Q,K)$ scale upon \emph{user demand} $U$ for using a service instance, as the ratio between the \emph{request rate} $R_{req}$ and the \emph{update frequency} $f_d$ \cite{HorvathKurt2023MESDD:Continuum} defined by the service operator:
\begin{equation}\label{eq8}
U = \frac{R_{\mathtt{req}}}{f_d}.
\end{equation}

We alter service state by satisfying four demand constraints: \mbox{$K=\left\{ \kappa_\mathtt{MIN}, \kappa_\mathtt{LOW}, \kappa_\mathtt{UP}, \kappa_\mathtt{MAX}\right\}$}.
 \paragraph{Underdemand}$\kappa_\mathtt{MIN}$ asserts a minimal service load:
 \begin{equation} \label{eq:knon}
 \kappa_{\mathtt{MIN}}(U) = \begin{cases} 
 \text{true}, & U > 0; \\
 \text{false}, & U = 0 \text{ (downscale)}.
 \end{cases}
 \end{equation}

 \paragraph{Low-demand} $\kappa_\mathtt{LOW}$ asserts if the current demand is still high enough or below the minimum threshold $U_{\min}$. To increase the stability of the demand constraint, we use a \emph{static hysteresis} $\sigma$ to delay the initial change \cite{MerriamWebsterDictionaryDefinitionHYSTERESIS}, generally applied on control systems in \cite{Branicky1998AnalyzingSystems}.
  \begin{equation} \label{eq:klow}
  \kappa_{\mathtt{LOW}}(U, \sigma) = \begin{cases} 
  \text{true}, & U + \sigma \leq U_{\min};\\
  \text{false}, & U + \sigma > U_{\min}. 
  \end{cases}
  \end{equation} 

 \paragraph{High-demand} $\kappa_\mathtt{UP}$ ensures that if the demand is higher than the minimum threshold $U_{\min}$ for the service instance to remain available.
 \begin{equation} \label{eq:kup}
 \kappa_{\mathtt{UP}}(U, \sigma) = \begin{cases} 
 \text{true}, & U - \sigma \geq U_{\min}; \\
 \text{false}, & U + \sigma < U_{\min}. 
 \end{cases}
 \end{equation} 

 \paragraph{Overdemand} $\kappa_\mathtt{MAX}$ evaluates if the current demand exceeds the maximum threshold per instance. This demand constraint does not lead to a transition but spawns a new instance when fulfilled to serve the increased demand:
 \begin{equation} \label{eq:kmax}
 \kappa_{\mathtt{MAX}}(U) = \begin{cases} 
 \text{true}, & U \geq U_{\max} \text{ (upscale)};\\
 \text{false}, & U < U_{\max}.
 \end{cases}
 \end{equation} 

Section \ref{sub:scaling} explicitly describes the scaling method.

\subsection{Service State Transitions}
There are six service state transitions depicted in Figure  \ref{fig:overviewServicestate-machine}:
\begin{equation}  \label{eq4}
\Sigma = \left\{T_\mathtt{DIS}, T_\mathtt{UND}, T_\mathtt{RED}, T_\mathtt{INA}, T_\mathtt{REA}, T_\mathtt{FIN} \right\}.
\end{equation} 

\paragraph{Discover} $T_\mathtt{DIS}$ transits a every service instance to a discoverable state $S_\mathtt{DIS}$ accessible to the user based on the underdemand constraint $\kappa_\mathtt{MIN}$ (Equation \ref{eq:knon}):
 \begin{equation} \label{tred}
 \begin{split}
T_\mathtt{DIS}(S_\mathtt{STO}, \kappa_\mathtt{MIN}) = S_\mathtt{DIS}.
\end{split}
\end{equation} 

\paragraph{Undiscover} $T_\mathtt{UND}$ transition makes the service unobtainable when the user demand declines below a minimum demand using the static hysteresis $\sigma$ to improve stability to avoid transitions on reaching $U_{\min}$ (Equation \ref{eq:klow}):
\begin{equation} \label{eq:tund}
T_\mathtt{UND}\left(S_\mathtt{DIS}, \kappa_\mathtt{LOW}\right)  = S_\mathtt{UND}.
\end{equation} 

\paragraph{Reinstate} $T_\mathtt{RED}$ transition makes the service available with high enough user demand $\kappa_{UP}$ (Equation \ref{eq:kup}):
\begin{equation} \label{eq:tred}
T_\mathtt{RED}(S_\mathtt{UND}, \kappa_\mathtt{UP})  = S_\mathtt{DIS}.
\vspace{-3mm}
\end{equation}.

\paragraph{Decomission} $T_\mathtt{INA}$ transitions disables the service for user requests upon an inactive service maintenance event $\alpha_\mathtt{INA}$  triggered by the service operator (Paragraph \ref{par:aIna}):
\begin{equation} \label{eq:tina}
T_\mathtt{INA}\left(S_\mathtt{UND},\alpha_\mathtt{INA}\right) = S_\mathtt{INA}.
\end{equation}
 
\paragraph{Reactivate} $T_\mathtt{REA}$ triggered by the service provider upon the maintenance event $\alpha_\mathtt{REA}$ acts as a gatekeeper to restrict availability (Paragraph \ref{par:Rea}):
\begin{equation} \label{eq:trea}
T_\mathtt{REA}\left(S_\mathtt{INA}, \alpha_\mathtt{REA}\right) = S_\mathtt{DIS}
\end{equation}

\paragraph{Finalize} $T_\mathtt{FIN}$ ends the service lifecycle in response to vanishing demand, reaching a final state $S_\mathtt{FIN}$ (Equation \ref{eq:knon}):
\begin{equation} \label{eq:tfin}
T_\mathtt{FIN}\left(S_\mathtt{INA},\kappa_\mathtt{MIN}\right) = S_\mathtt{FIN}
\end{equation}

\section{Scaling Methodology}\label{sec:problem}
Employing elastic scaling for service placement is essential for efficient resource usage and meeting user demand. This section discusses how the FSM organizes the life cycle of a service and extends it further by an indicator to prevent preemptive transitions. Furthermore, the complete algorithm is delineated.

\subsection{Temporal Stability}\label{sub:temporalstability}
We define the \textit{temporal stability indicator} $I_R$ to evaluate requests during an observation period $\Delta t$ as:
    \begin{equation}\label{eq:transitionratecummulate}
    I_R = \left|\frac{k}{\Delta t}\right|,
    \end{equation}
where $k$ counts the number of service requests on a node during the observation period $\Delta t$, which must be much higher than the unit of the request rate (e.g., $R_\mathtt{req}$ = $\qty{10}{\second}, \Delta t = \qty{10}{\minute}$) to ensure that the lifecycle management does not remove service instances due to short-lived fluctuations in demand. However, if the user demand increases fast, the scaling method ensures service availability or even deploys new service instances based on three conditions:
\paragraph{$|R_\mathtt{req} = I_R|$} indicates a temporal stable deployment.
\paragraph{$R_\mathtt{req} \ll I_R$} indicates am oversized deployment to downscale.
\paragraph{$R_\mathtt{req} \gg I_R$} indicates a much higher incoming request rate requiring upscaling.

To evaluate all three conditions, we define lower and upper bounds for the stability indicator:
$0 \leq I_R^{\min} < I_R < I_R^{\max}$.

\textit{Example}: We asses the temporal stability of a node and define $I_R^{\min} = 10$ and $I_R^{\max} = 250$. If $R_\mathtt{req} = 350$, there is no immediate need to change the service state. If $R_\mathtt{req}$ remains longer at a higher value, ${I_R}$ also increases since it has the same input but a longer observation period and triggers a transition.
However, if frequency $f_d=15$, then $U_{\max} = \frac{300}{15}=20$ (Equation \ref{eq:kmax}), $\kappa_\mathtt{MAX}$ returns true and scaling applies.

\subsection{Service Scaling Algorithm}\label{sub:scaling}
We perform the scaling process considering the demand constraints described in Section \ref{sub:demandbasedtrans}, as represented in Algorithm \ref{alg:scaling}.
We define $R_{req}$, $f_d$, and $\Delta t$ as input parameters to the algorithm (Lines 2–4). The algorithm's output is $S'$, representing the FSM's next state (Line 6). Next, we initialize three variables ($k, n, t$) used to calculate $I_R$; these variables operate globally.

\begin{algorithm}[t]
\caption{SCAREY service scaling algorithm.}\label{alg:scaling}
   
  \begin{algorithmic}[1]
  \Input
  \Desc{$R_{\mathtt{req}}$} \Comment{request rate}
  \Desc{$f_d$} \Comment{update frequency}
  \Desc{$\Delta t$} \Comment{observation period}
  \EndInput

 \item []
\Parameter{global}{$k \gets 0$} \Comment{initialize temporal stability}
\Parameter{global}{$n \gets 0$} 
\Parameter{global}{$t \gets \Call{CurrentTime()}{}$} 
 \item []
\Procedure{ScaleService}{$R_{\mathtt{req}}$, $f_d$, $\Delta t$}
\State $S' \gets S_{\mathtt{STO}}$ \Comment{initial state}
\State $I_R \gets {R_{\mathtt{req}}}$ 

\While{$S' \neq S_{\mathtt{FIN}}$}
 \State $U \gets \frac{R_{\mathtt{req}}}{f_d}$
 \State $I_R \gets \Call{CalcTempStability}{R_{\mathtt{req}}, \Delta t, I_R}$
  \State $S' \gets \Call{UpdateService}{U, I_R, S'}$  \Comment{update FSM}
\EndWhile
\EndProcedure
 \item []
\Function{CalcTempStability}{$R_{\mathtt{req}}$, $\Delta t$, $I_R$}
\State $N \gets  \Call{CurrentTime()}{} $ \Comment{acquire timestamp}
\If{ $N \leq t + \Delta t$} \Comment{cummulate $k$ until $\Delta t$ }
\State $k \gets k + R_{\mathtt{req}}$
\State $n \gets n + 1$
\State \Return $I_R$
\Else
 \State $I_R \gets \frac{k}{n}$ \Comment{$\Delta t$ reached}
 \State $k \gets 0, n \gets 0$
\State $t \gets \Call{CurrentTime()}{} $
\State \Return $I_R$
\EndIf
\EndFunction
 \item []

 \Function{UpdateService}{$U$, $I_R$, $S'$}
\If{$\kappa_{\mathtt{MIN}}(U) = \mathbf{False} \lor I_R < I_R^{\min}$} 
    \State $S' \gets S_{\mathtt{FIN}}$  \Comment{scale-down}


\ElsIf{$\kappa_{\mathtt{UP}}(U) = \mathbf{True} \land \kappa_{\mathtt{MAX}}(U) = \mathbf{False}$}
    \State $S' \gets S_{\mathtt{DIS}}$

\ElsIf{$\kappa_{\mathtt{MAX}}(U) = \mathbf{True} \lor I_R \geq I_R^{\max}$} 
    \State $S' \gets S_{\mathtt{DIS}}$  \Comment{scale-up}
    \State $\Call{createFSM}$
\Else
 \State $S' \gets S_{\mathtt{UND}}$  \Comment{low demand}
\EndIf
\State \Return $S'$
\EndFunction

\end{algorithmic}
\end{algorithm}

The algorithm begins with the \texttt{ScaleService} procedure (Line 7). Next, it sets $S'$ to the stored state $S_{STO}$ (Line 6) and assigns the current request rate to the temporal stability indicator $I_R$.
In the next step, we iteratively update the state machine until it reaches its final state $S_{FIN}$ (Line 8).
To determine the current user demand $U$, as described in Section \ref{sub:demandbasedtrans}, we divide the request rate $R_{req}$ by the update frequency $f_d$ (Line 9). To ensure deployment stability, the algorithm calculates the temporal stability indicator $I_R$ using the function \texttt{CalcTempStability($R_{req}$, $\Delta t$, $I_R$)} (Line 10).
The next step evaluates whether a new state is applicable based on the updated values of $U$ and $I_R$ and the current state $S'$. This evaluation is performed in Line 11 using the function \texttt{UpdateService($U$, $I_R$, $S'$)}.

The function \texttt{CalcTempStability($R_{req}$, $\Delta t$, $I_R$)}, defined in Line 14, follows the steps outlined in Section \ref{sub:temporalstability}. First, it retrieves the current time (Line 15) and checks whether the observation period $\Delta t$ has elapsed (Line 16). If $\Delta t$ has not yet passed, the algorithm accumulates the number of requests $R_{req}$ in $k$ and increments the count $n$ (Lines 17–18) and returns the previous value of $I_R$ since it is still valid. After the passage of $\Delta t$, the system resets updates $I_R$, resets $k$, $n$ to zero, and sets $t$ to the current time for the next iteration (Lines 22-23). Finally, the function returns the updated value of $I_R$ (Line 24). 

The \texttt{UpdateService($U$, $I_R$, $S'$)} function evaluates conditional statements to determine whether the current state $S'$ needs to be modified.
The FSM terminates if $U = 0$ and the condition $\kappa_{min}(U)$ is satisfied (Line 28). If $I_R$ does not exceed the minimum stability threshold $I_R^{min}(n)$, the state transitions to $S_{FIN}$, allowing the algorithm to exit (Line 30) the while-loop in the \texttt{ScaleService} procedure.
To determine if the service instance can become discoverable ($S_{DIS}$), the algorithm evaluates the current user demand $U$ against the thresholds $\kappa_{MAX}(U)$ and $\kappa_{UP}(U)$ (Lines 30-31). When the algorithm reaches $\kappa_{MAX}(U)$ or if temporal stability meets $I_R^{max}(n)$, it creates a new service instance.


\section{SCAREY Architecture}\label{sec:framework}

This section presents a detailed architecture of the SCAREY approach illustrated in Figure \ref{fig:implementation} and consisting of four modules:
\begin{enumerate*}
\item{Lifecycle management core} for executing the service scaling algorithm and updating service instances as needed.
\item{Infrastructure} for hosting and provisioning of the service instances.
\item{Geographic management} for ensuring the validity of the nodes assigned to a user.
\item{Service discovery} allowing users to locate and access the service instances.
\end{enumerate*}

 \vspace{-2mm}
\subsection{User and Service Discovery} \label{sub:serviceDiscovery}
This module implements Mobile Edge Service Discovery using DNS (MESDD)\cite{HorvathKurt2023MESDD:Continuum}.
The user device runs the MESDD discovery process to find the best service instance in the infrastructure with the lowest latency. MESDD serves as a user-oriented service discovery scheme, using a lookup to determine the user's location. The system organizes services and users into distinct geographical zones, linking them to nodes. The DNS server administers nodes based on the resolved zone names.
 \vspace{-2mm}
\subsection{Geographic Management}\label{sub:geoContext}
This module provides a persistence layer that supports the dynamic scheduler and deployment manager in identifying and deploying services efficiently. It determines a user's and service locations through a location lookup process, enabling users to retrieve services based on their geographical position. 
The geographic manager application maintains (I) the link between zones and nodes to enhance service discovery, optimizing performance and accessibility.
 \vspace{-2mm}
\subsection{Lifecycle Management Core}\label{subsec:coreModule}
The lifecycle management core manages the deployment of service instances based on the users' demand following the lifecycle state machine, considering the service scaling algorithm proposed in Section \ref{sub:scaling}. It encompasses the following three applications.

\begin{figure}[t]
    \centering
        \vspace{-4mm}
    \includegraphics[width=0.45\textwidth]{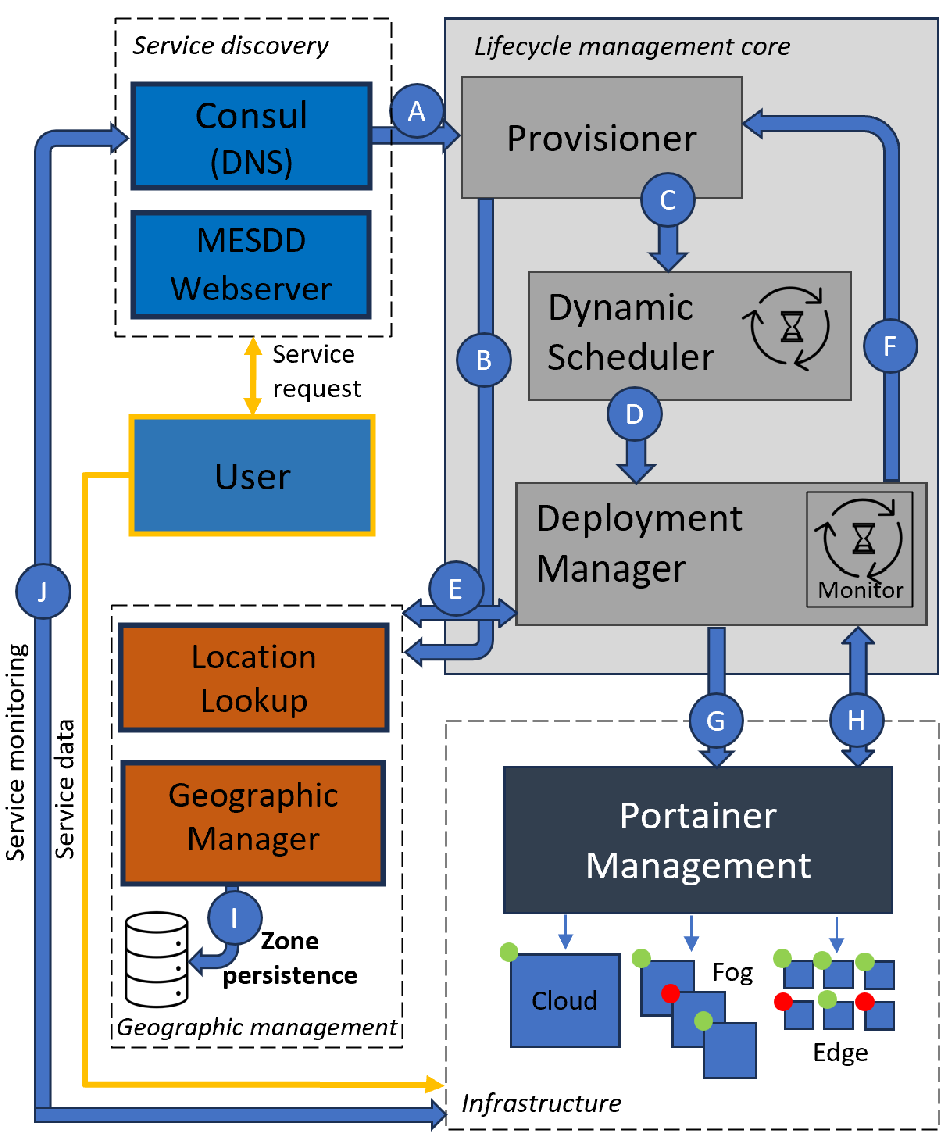} 
    \caption{SCAREY Architecture}%
    \label{fig:implementation}%
        \vspace{-6mm}
\end{figure}

\paragraph{Provisioner} receives each request for a service instance from the service discovery module (A). Then, it validates if the node is available using the geographic management module (B). Next, the provisioner forwards the request to the dynamic scheduler (C).

\paragraph{Dynamic scheduler} \label{par:statcore} handles all incoming requests forwarded by the provisioner. For each service instance, it aggregates requests according to the service scaling algorithm \ref{alg:scaling} detailed in Section \ref{sub:scaling}, applying state machine transitions for each instance. As user demand evolves, the state machine for each service instance is updated accordingly. During scaling operations, the scheduler launches a new service instance on a different node or initiates another one on the same node and informs the deployment manager (D) of changes for provisioning the nodes and service instances.

\paragraph{Deployment manager} \label{par:deploymentmgr} performs the deployment (G) of the services based on the plan provided by the dynamic scheduler, as illustrated in Figure \ref{fig:implementation}, supported by a monitoring tool, which observes the current state of the infrastructure (H) and reports possible unresolvable constraints (F). For example, if a hardware failure prevents a service instance from being deployed, the provisioner can discard the requests, which keeps unnecessary load off the dynamic scheduler. A node could become unavailable for operations due to maintenance. Geographic management (E) confirms that the node and its location assigned to it are unavailable.

\subsection{Infrastructure} \label{sub:containerManagement}
The infrastructure module contains portainer management, a lightweight service delivery platform for containerized applications\footnote{\url{https://www.portainer.io/}}, providing access to the infrastructure available for deployment. This infrastructure reflects the computing continuum encompassing hosts across Cloud, Fog, and Edge domains. The portainer management allows other modules, like the lifecycle management core, to alter the current deployment using a REST API. Finally, in case of successful service discovery, the user device can access the infrastructure by addressing available nodes. 

\section{Evaluation}\label{sec:kpi}
This section outlines the testbed design, including the deployment environment and configurations, followed by a reference application and evaluation metrics. Finally, the section defines real-world and simulated scenarios. 

\subsection{Testbed}\label{sub:testbed}
We deploy compute nodes across the CC, considering three layers, Cloud, Fog, and Edge, summarized in Table \ref{tab:comparison}.


\begin{table}[t]

\resizebox{\linewidth}{!}{%
\begin{tabular}{|@{}l@{}|@{}c@{}|@{}c@{}|@{}c@{}|@{}c@{}|}
\hline
\textbf{Feature} & \textbf{\begin{tabular}[c]{@{}c@{}}NVIDIA \\ Jetson Nano\end{tabular}} & \textbf{\begin{tabular}[c]{@{}c@{}}AWS \\ EC2 t2.small\end{tabular}} & \textbf{\begin{tabular}[c]{@{}c@{}}AWS \\ EC2 t2.xlarge\end{tabular}} & \textbf{\begin{tabular}[c]{@{}c@{}}AWS \\ EC2 t4g.2xlarge\end{tabular}} \\ \hline
\textbf{Use} & Edge Layer & Fog Layer & Cloud Layer & Cloud Layer \\ \hline
\textbf{CPU} & \begin{tabular}[c]{@{}c@{}}Quad-core ARM \\ Cortex-A57\end{tabular} & \begin{tabular}[c]{@{}c@{}}1 vCPU \\ (Intel Xeon)\end{tabular} & \begin{tabular}[c]{@{}c@{}}4 vCPU \\ (Intel Xeon)\end{tabular} & \begin{tabular}[c]{@{}c@{}}8 vCPU \\ (Intel Xeon)\end{tabular} \\ \hline
\textbf{GPU} & \begin{tabular}[c]{@{}c@{}}128-core \\ NVIDIA Maxwell\end{tabular} & N/A & N/A & N/A \\ \hline
\textbf{Memory} & 4 GB LPDDR4 & 2 GB DDR3 & 16 GB DDR3 & 32 GB DDR3 \\ \hline
\textbf{Storage} & microSD (16 GB) & EBS (80 GB) & EBS (80 GB) & EBS (80 GB) \\ \hline
\textbf{Network} & Gigabit Ethernet & \begin{tabular}[c]{@{}c@{}}Up to 1 Gigabit \\ (shared)\end{tabular} & \begin{tabular}[c]{@{}c@{}}Up to 1 Gigabit \\ (burst)\end{tabular} & \begin{tabular}[c]{@{}c@{}}Up to 5 Gigabit \\ (burst)\end{tabular} \\ \hline
\textbf{\begin{tabular}[c]{@{}l@{}}Operating\\System\end{tabular}} & Ubuntu Linux & Amazon Linux & Amazon Linux & Amazon Linux \\ \hline
\textbf{\begin{tabular}[c]{@{}l@{}}Power \\Consumption\\on CPU load\end{tabular}} & \begin{tabular}[c]{@{}c@{}}2.4 W (idle) \\ 4.1 W (10\%)\\ 6.8 W (50\%)\\ 8.8 W (100\%)\end{tabular} & \begin{tabular}[c]{@{}c@{}}2.0 W (idle) \\ 3.3 W (10\%)\\ 5.3 W (50\%)\\ 7.0 W (100\%)\end{tabular} & \begin{tabular}[c]{@{}c@{}}9.6 W (idle) \\ 15.7 W (10\%)\\ 24.6 W (50\%)\\ 33.0 W (100\%)\end{tabular} & \begin{tabular}[c]{@{}c@{}}12.3 W (idle) \\ 19.3 W (10\%)\\ 30.7 W (50\%)\\ 42.0 W (100\%)\end{tabular} \\ \hline
\textbf{\begin{tabular}[c]{@{}l@{}}CO2 Emissions\\ on CPU load\end{tabular}} & \begin{tabular}[c]{@{}c@{}}1.0 g (idle) \\ 1.7 g (10\%)\\ 2.9 g (50\%)\\ 3.7 g (100\%)\end{tabular} & \begin{tabular}[c]{@{}c@{}}0.8 g (idle) \\ 1.4 g (10\%)\\ 2.2 g (50\%)\\ 2.9 g (100\%)\end{tabular} & \begin{tabular}[c]{@{}c@{}}4 g (idle) \\ 6.6 g (10\%)\\ 10.4 g (50\%)\\ 13.9 g (100\%)\end{tabular} & \begin{tabular}[c]{@{}c@{}}5.2 g (idle) \\ 8.1 g (10\%)\\ 12.9 g (50\%)\\ 17.7 g (100\%)\end{tabular} \\ \hline
\textbf{Cost per hour} & \$0.0614 & \$0.023 & \$0.185 & \$0.2688 \\ \hline
\end{tabular}
}
\caption{Available resources}

        \vspace{-4mm}
\label{tab:comparison}
\end{table}

\paragraph{Edge layer} nodes are closest to the user and are accessible with the lowest latency. We use Nvidia Jetson Nanos with limited CPU and GPU computational power but consume very little electrical power.
This level contains several instances used by our reference application, clustered in zones, as defined in Section \ref{sub:topology}.
Each zone aligns with the daily movements of its inhabitants. For example, users reside at night [19:00-07:00] hour in the \texttt{city center} and transit to the corporate area to commercial \texttt{east} from [08:00-18:00] hour. We consider the location of the inhabitants to define a total of eight zones (see Table \ref{tab:timewindowtopological}). 

\paragraph{Fog layer}
provides AWS EC2 computing nodes placed beyond the urban area. 
The nodes deliver services to a broader area and connect the zones covered in the Edge layer. Nodes in the Fog layer
 still ensure adequately low latency. When demand fluctuates and services scale down, we remove service instances, and users migrate to the Fog layer instances to maintain service continuity with moderate latency.

\paragraph{Cloud layer}
holds the most AWS EC2 computational power and provides service discovery to identify the best instance for a specific user. The Cloud layer also ensures the orchestration of nodes and service instances, operated by the \texttt{Statistical Core} (see Paragraph \ref{par:statcore} and the \texttt{Deployment Manager} \ref{par:deploymentmgr}).

\paragraph{Topology}\label{sub:topology}
defines eight zones based on temporal visitation patterns defined in Table \ref{tab:timewindowtopological}, as illustrated in Figure \ref{fig:scenarioover}. Within each defined zone, users engage with services at specific locations or continuously throughout the day. These zoning strategies enable the planning of service demand within the Edge layer. Our scenario does not divide the Fog and Cloud layers into zones, but statically provisions the resources. 
\begin{figure}[t]
    \centering
        \vspace{-2mm}
    \includegraphics[width=0.45\textwidth]{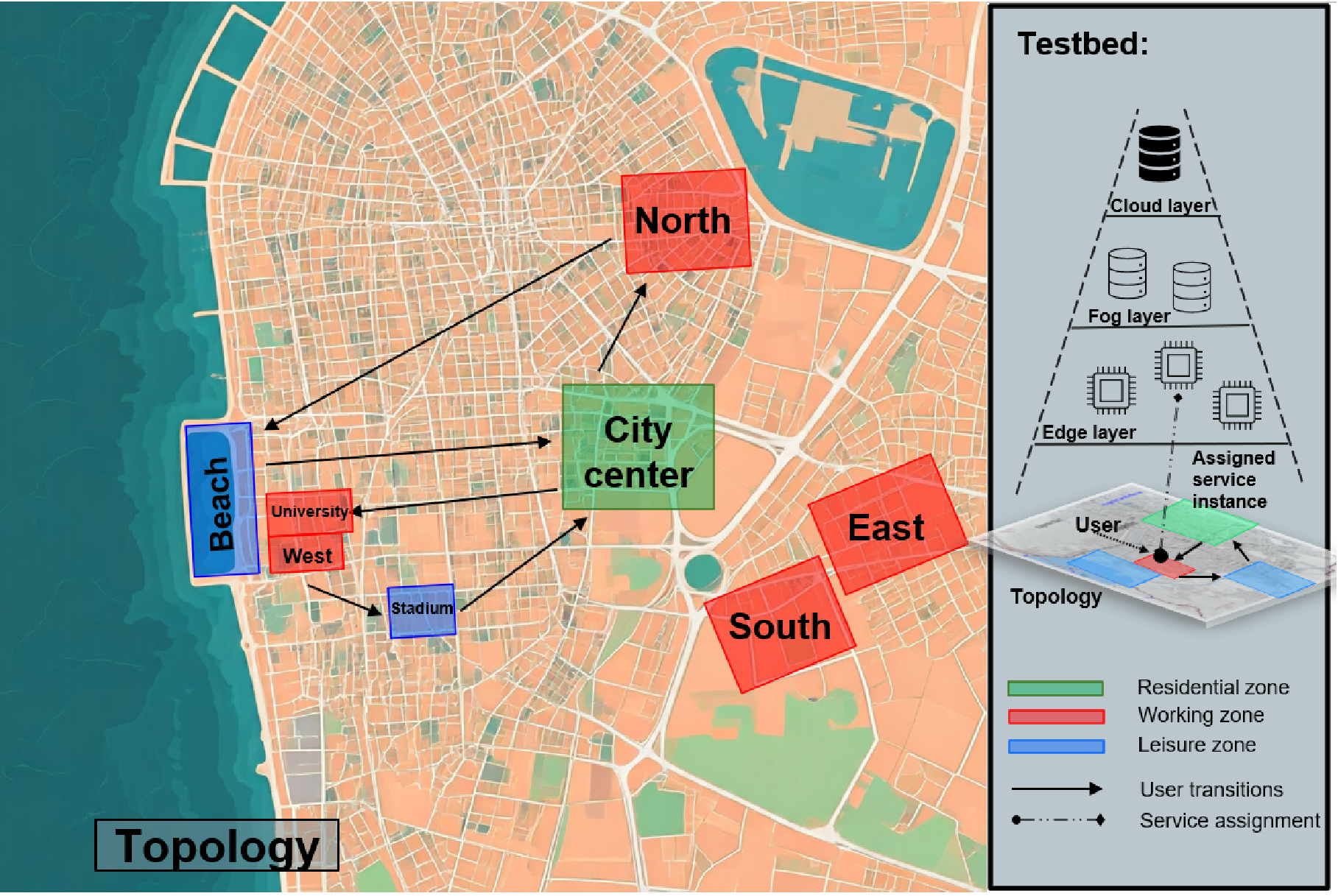} %
    \caption{Overview Edge layer NNC}%
    \label{fig:scenarioover}%
        \vspace{-2mm}
\end{figure}

\begin{table}[t]
\vspace{+0mm}
\centering
\scalebox{0.9}{
\begin{tabular}{|l|c|l|}
\hline
\textbf{Name} & \multicolumn{1}{l|}{\textbf{Time window}} & \textbf{Category} \\ \hline
City center & 19:00-07:00 & residential \\ \hline
Commercial North/South/East/West & 08:00-18:00 & working \\ \hline
University campus & 08:00-20:00 & working \\ \hline
Stadium & 18:00-22:00 & leisure \\ \hline
Beach & 09:00-19:00 & leisure \\ \hline
\end{tabular}
}
\caption{Time windows on topological zones}
\label{tab:timewindowtopological}
    \vspace{-4mm}
\end{table}

\subsection{ARlive Reference Application}\label{sub:refApp}
We conceptualize an application called \texttt{ARlive} using Augmented Reality (AR) location-based interaction items to deliver an immersive social shopping experience augmented with gamification elements\cite{Roddiger2018ARMart:Items}.

 \begin{figure}[!t]
     \centering
         \vspace{-2mm}
     \includegraphics[width=0.4\textwidth]{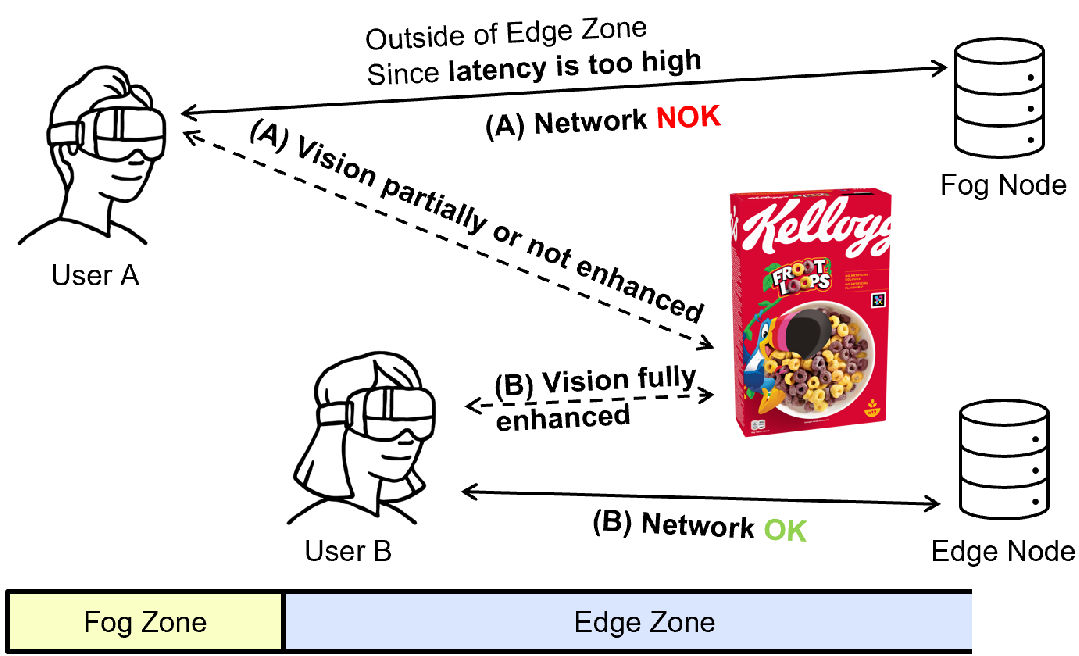} %
     \caption{ARlive use case schematic}%
     \label{fig:arlive}%
         \vspace{-4mm}
 \end{figure}

The design objective is to superimpose virtual objects and information onto the physical \cite{Ventoulis2022ARMythology} world via users' smartphones or AR glasses, enhancing the user's vision. 
The application alerts users to various interactive challenges and tasks associated with nearby stores and products. For example, the application may prompt a user to locate specific items within a store, scan QR codes, or participate in mini-games individually or collaboratively. 
As depicted in Figure \ref{fig:arlive}, the processing node for a user depends on her location and service provider. User \texttt{A} is outside an Edge zone and currently cannot meet the low latency requirements. 
Subsequently, AR features do not enhance her vision \cite{Syed2022In-depthConcerns}, where user \texttt{B} resides in an Edge zone and can fulfill the latency requirements. The vision of user \texttt{A} will be enhanced as long she remains in an Edge zone.
Figure \ref{fig:arliveimpl} depicts a real-world implementation of this scenario inside of a supermarket.

 \begin{figure}[t]
     \centering
         \vspace{-2mm}
     \includegraphics[width=0.5\textwidth]{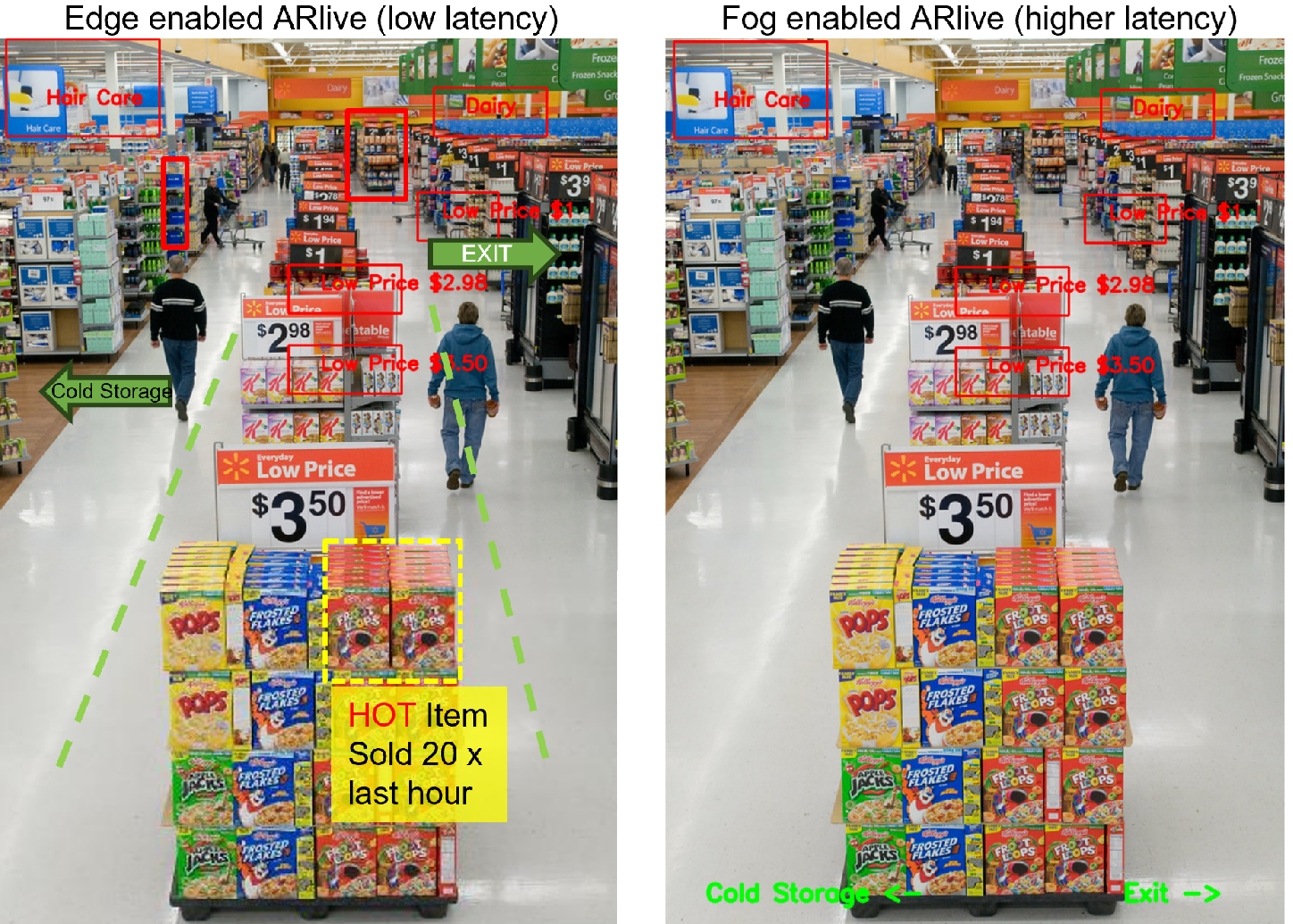} %
     \caption{ARlive implementation}%
     \label{fig:arliveimpl}%
         \vspace{-2mm}
 \end{figure}

The left side of Figure \ref{fig:arliveimpl} shows a complete set of AR features accessible to the user, highlighting offers, guidance projected to the ground and the popularity of items. This is possible since the low latency allows more processing, which still provides real-time information. On the other hand, the right side of the image shows a reduced set of information since the communication leaves less time for the processing to provide real-time enhancements.

\subsection{Metrics}\label{sub:metrics}
For evaluation purposes, we define the following metrics relevant to the users and the service providers. The first two evaluate the performance over time. The last three metrics assess financial and sustainability aspects by considering energy consumption and $CO_2$ emissions based on parameters defined in Table \ref{tab:comparison}.


\paragraph{Service discovery} is the time required to identify the node to service a client, including the time to inform the client of the service instance URL.
     
\paragraph{Service acquisition} is the time required from the occurrence of the demand until the user receives the service data from the assigned node.


\paragraph{Cost} for the used resources computing continuum resources in the infrastructure.
    \begin{equation}
    C_{\text{total}} = \sum_{\text{type} \in \{\text{edge, fog, cloud}\}} \sum_{i=1}^{n_{\text{type}}} ( t_i \cdot R_{\text{type}})
    \end{equation}
    where:
    \begin{itemize}
        \item $t_i$ is the time for using an instance instance $i$;
        \item $R_{\text{type}}$ is the cost per time unit for the instance type (edge, fog, or cloud) as defined in Table \ref{tab:scenarios}.
        \item $n_{\text{type}}$ is the number of instances of the given type.
    \end{itemize}

   \paragraph{Carbon emissions} for executing the service across the computing continuum.
    \begin{equation}
    E_{\text{total}} = \sum_{\text{type} \in \{\text{edge, fog, cloud}\}} \sum_{i=1}^{n_{\text{type}}} (t_i \cdot E_{\text{type}})
    \end{equation} 
    where: $E_{\text{type}}$ are the $CO_2$ emissions in grammes per hour per instance type.

    \paragraph{Power consumption} for using the entire computing continuum instances.
    \begin{equation}
    P_{\text{total}} = \sum_{\text{type} \in \{\text{edge, fog, cloud}\}} \sum_{i=1}^{n_{\text{type}}} (t_i \cdot P_{\text{type}})
    \end{equation}
    where $P_{\text{type}}$ is the power consumption rate (energy per unit time) for the instance type (edge, fog, or cloud).
For the power consumption and carbon emissions of the infrastructure, we use estimations as defined in \cite{Mytton2022SourcesReview}, which are based on the dataset of Davy \cite{BenjaminDavy2021BuildingDataset}. 


\subsection{Related Approaches}\label{sub:relatedapproach}
To evaluate our approach, we selected two related methods described in Section \ref{sec:relatedWork}, namely COLAP \cite{Gupta2017COLAP:Environment} and PIES \cite{Hudson2021QoS-AwareImplementations}. The two approaches leverage REST APIs to enable deployment on the testbed, dynamic interactions and real-time adaptability. These modifications ensure compatibility with our setup while maintaining their core functionality.

\subsection{Evaluation Scenarios}\label{sub:scenario}
We selected three scenarios to evaluate the scalability features of SCAREY. 

\paragraph{Scale-up from low demand}\label{par:scaleup} This scenario starts with a minimal user demand as defined in Equation \ref{eq8}. All service instances within the \textit{Edge layer} start in an inactive state, represented as $S_{INA}$. Initially, user demand increases within our testbed's \texttt{city centre} between [19:00-07:00] hours (see Table \ref{tab:timewindowtopological}. Users transit to working zones (\texttt{commercial north/south/east/west}) with a velocity of \SI{15}{\meter\per\second} and reside in the time window of [08:00-18:00] on average for 8.5 hours. Users from \texttt{commercial south} and \texttt{West} will transit to the \texttt{Stadium} with a speed of \SI{10}{\meter\per\second} and reside for two hours before they finally transit to the \texttt{city centre} with a velocity of \SI{10}{\meter\per\second} and reside for 12.5 hours. Where the other 50\% of users, after work, directly transit back to the \texttt{city center} with a velocity of \SI{15}{\meter\per\second}. 

\paragraph{Scale-down from high demand}\label{par:scaleDown}
 This scenario begins with a high user demand $U$ (peak load, defined in Table \ref{tab:scenarios}), where all service instances within the \textit{Edge layer} operate in a discoverable state, represented as $ S_{DIS} $. User transits are identical with the scenario 'scale-up from low demand'. However, user demand decreases since users finish the defined runs of the user transits. The objective is to demonstrate how service instances reduce their operations during the latter half of the day in response to the declining user request loads.
 
\paragraph{Underprovisioning of computing nodes}\label{par:scenUnderprov}
In this scenario, we build upon the increase in user demand $U$ discussed in the 'scale-up from low demand'-scenario, doubling the number of user requests after \SI{10}{\minute}. Simultaneously, the number of active zones is reduced until all users interact with a single node. This configuration pushes the load toward the hardware's capacity limits, potentially positioning the Fog node as the primary source of service delivery.

\begin{table}[t]
\centering
\scalebox{0.85}{
\begin{tabular}{|@{}>{\raggedright\arraybackslash}p{1.99cm}@{}|>{\raggedright\arraybackslash}p{1.4cm}|>{\raggedright\arraybackslash}p{1.4cm}|>{\raggedright\arraybackslash}p{2.0cm}|>{\raggedright\arraybackslash}p{2.0cm}|}
\hline
\emph{Parameter} & \emph{Scale-up from low demand} & \emph{Scale-down from high demand} & \emph{Underprovision computing nodes} & \emph{Annual user demand} \\
\hline
Initial requests & 0 & Peak load (200 active users) & 0 & 0\\
\hline
Active users & 0 & Peak load (200 active users) & 0 & 1000 \\
\hline
User increment & 5 per minute & Decrease by 5 per minute & Double user requests after 10 minutes & 5 per minute \\
\hline
User discovery interval & 600 seconds & 600 seconds & 600 seconds & 600 seconds \\
\hline
User update interval & 5 seconds & 5 seconds & 5 seconds & 5 seconds \\
\hline
Peak load & 200 active users & 200 active users & All user requests on the Fog node & 1000 active users\\
\hline
Hold peak load & 10 minutes & 10 minutes & Until hardware limitations occur or the load can be processed for 10 minutes & From [19:00 - 05:00], declines by five users per minute until 0; From [05:00] increases again with the given user increment \\
\hline
\end{tabular}
}
\caption{List of the parameters for the defined scenarios}
\label{tab:scenarios}
\vspace{-2mm}
\end{table}

\paragraph{Annual user demand}\label{par:scenAnnualdemand}
This scenario extrapolates on the scale-up and scale-down scenarios based on the time windows defined in Table \ref{tab:timewindowtopological}. Users traverse in the morning from the \texttt{city center} to commercial zones and at [19:00] back to the \texttt{city center}. This pattern is repeated every day of the year and provides the time durations to define the expected user demand $U$ and the subsequent resource demand. This scenario is applied to calculate annual costs, the annual $CO_2$ emissions and the annual electricity costs to operate a deployment for 1000 users.

\par
Table \ref{tab:scenarios} describes the variable parameters used in each scenario. Each zone defines the time window when a service host must be available. Since SCAREY identifies the correct service instance based on the user's location, it switches off inaccessible nodes (instance reach state: $S_{FIN}$). This indicates that eight nodes positioned for eight zones have an effective operational time of \SI{9.75}{\hour\per\day} which is applied in the scenario 'annual user demand' found in Paragraph \ref{par:scenAnnualdemand}.

\begin{figure*}[ht]
    \centering
    \subfloat[\centering COLAP]{{\includegraphics[width=0.32\textwidth]{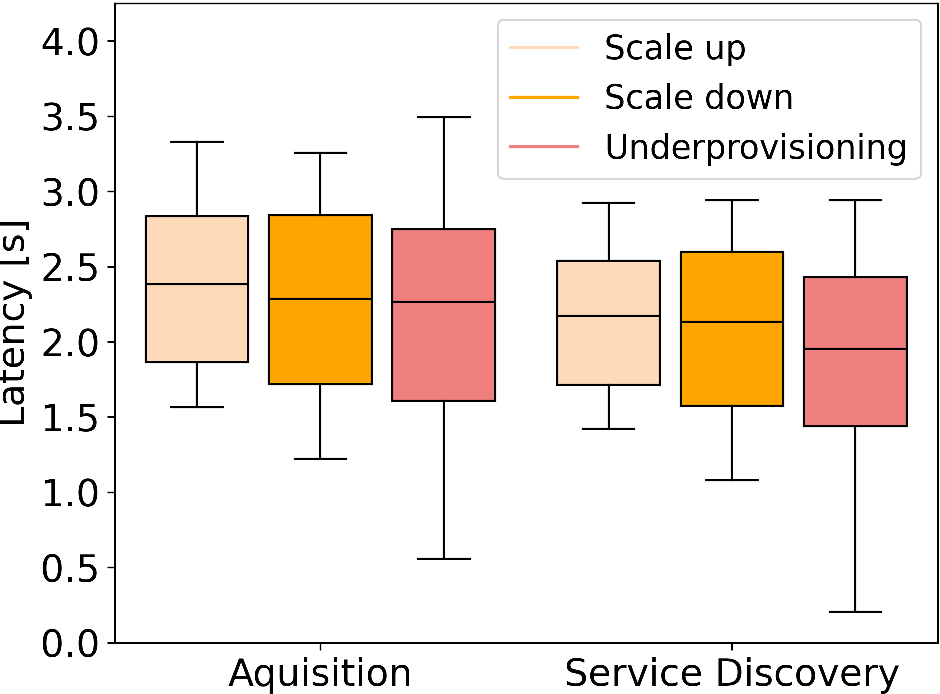} }\label{fig:performanceColap}}%
    \
    \subfloat[\centering PIES]{{\includegraphics[width=0.32\textwidth]{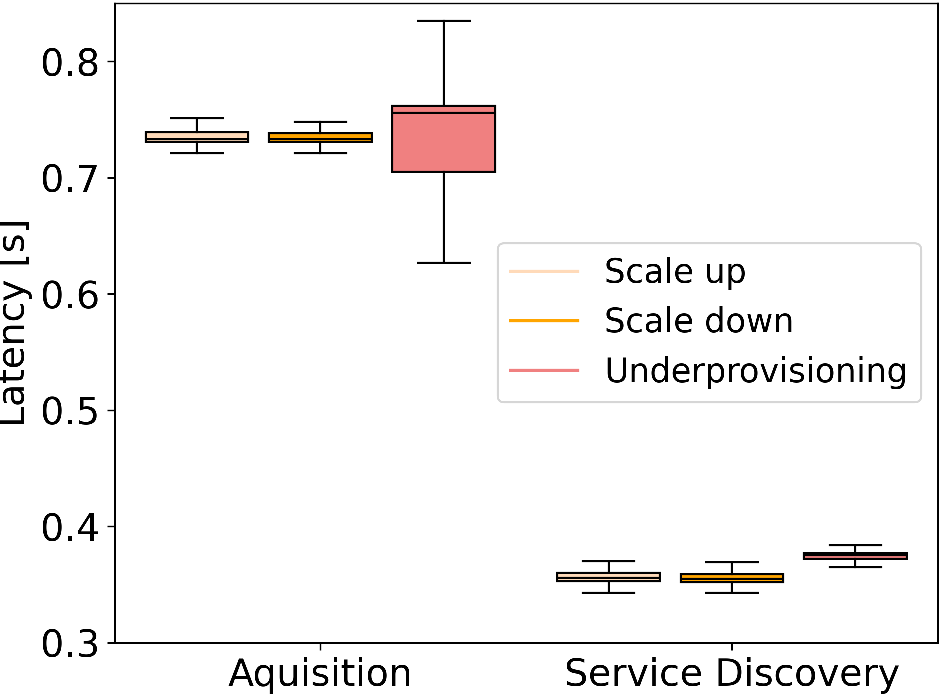} }\label{fig:performancePies}}%
    \
    \subfloat[\centering SCAREY]{{\includegraphics[width=0.32\textwidth]{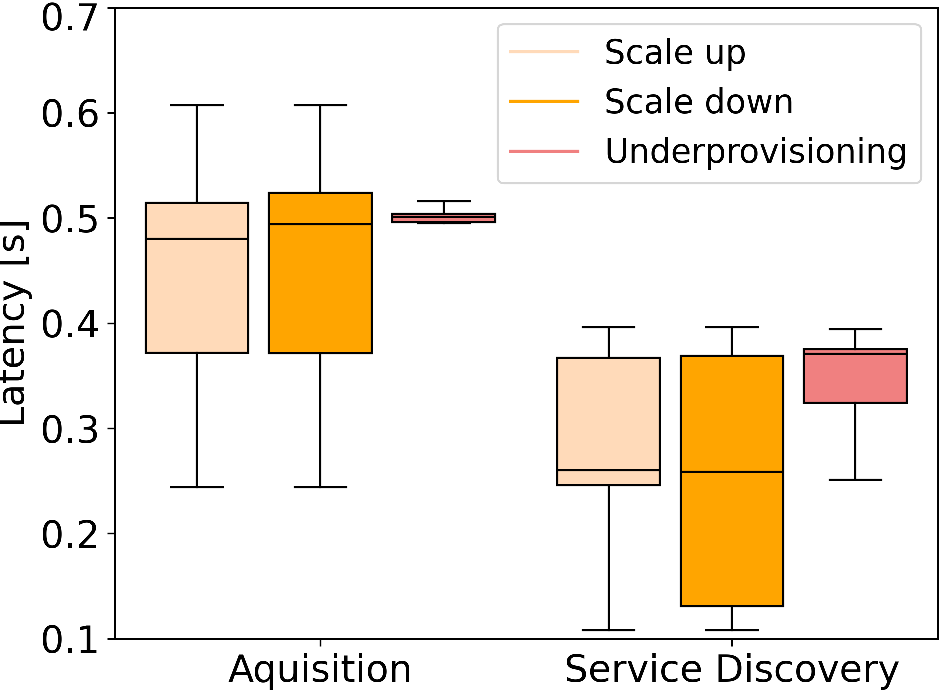} }\label{fig:performanceScarey}}%
    \label{fig:performance}%
    \caption{performance related methods and SCAREY for service discovery and service acquisition.}%
    \vspace{-4mm}
\end{figure*}

\section{Results}\label{sec:Results}
The following section evaluates the performance of SCAREY in comparison with two service management approaches using five metrics: service discovery time, service acquisition time, annual costs, $CO_2$ emissions, and power consumption.

\subsection{Performance COLAP}\label{sub:COLAP} Figure \ref{fig:performanceColap} illustrates COLAP's service discovery and acquisition performance for the three scenarios: scale-up, scale-down, and underprovisioning.
COLAP consistently exhibits the longest mean discovery time for service discovery, with \SI{2135}{\milli\second} recorded in the scale-up scenario. This trend persists across all scenarios, highlighting its inefficiency in dynamic environments. The method struggles to adapt quickly to changing conditions, resulting in prolonged discovery times.
COLAP is the slowest method for service acquisition, recording a mean acquisition time of \SI{2366}{\milli\second}. Its high acquisition time underscores its inefficiency in discovering and acquiring services, making it the least siutable approach for latency-sensitive applications.

\subsection{Performance PIES}\label{sub:PIES} Figure \ref{fig:performancePies} presents the results for PIES across the same scenarios.
PIES achieves the shortest mean discovery time across all scenarios for service discovery, with \SI{356}{\milli\second} recorded in the scale-up scenario. PIES maintains stable performance with minimal variability, ensuring efficient service discovery even under changing network conditions.
Regarding service acquisition, PIES has a mean acquisition time of \SI{740}{\milli\second}, significantly lower than COLAP but slightly higher than SCAREY. The acquisition performance remains consistent across different scenarios, demonstrating PIES' ability to maintain reliable response times without significant fluctuations.

\subsection{Performance SCAREY}\label{sub:SCAREY} Figure \ref{fig:performanceScarey} further illustrates the performance of our approach SCAREY for the same scenarios.
For service discovery, SCAREY achieves a mean discovery time of \SI{255}{\milli\second} in the scale-up scenario, placing it between PIES and COLAP. However, SCAREY exhibits a higher standard deviation, indicating more significant variability in discovery times. The complexity of the MESDD-based discovery process results in this variability, as it requires additional steps compared to PIES. Despite this, SCAREY improves its stability in the underprovisioning scenario, where the reduced number of available nodes simplifies load balancing.
For service acquisition, SCAREY achieves the shortest mean acquisition time bellow \SI{500}{\milli\second}, outperforming both PIES and COLAP. Its advantage lies in its MESDD discovery scheme, which enables optimal node selection with the lowest possible latency.


\subsection{Service Costs}\label{sub:costs}
Figure \ref{fig:costs} depicts the annual service costs for the three different approaches based on the annual user demand defined in Paragraph \ref{par:scenAnnualdemand}. SCAREY induces a total annual cost ($C_{total}$) of \$3,791, with \$1,630 attributed to the Cloud, \$406 to the Fog, and \$1,754 to the Edge. In contrast, COLAP and PIES incur similar annual costs for Fog and Edge with \$201 and \$4318 but show increased total costs of \$6842 due to increased performance demand in the Cloud. SCAREY achieves its cost advantage by utilizing the MESDD strategy, which identifies Edge instances without active users and enables their suspension, thereby reducing unnecessary expenses. Furthermore, in the scenario, each node is active for an average of 9.75 hours per day, enabling SCAREY to optimize resource usage and minimize costs by adjusting service availability based on actual demand, resulting in a cost reduction of 45\%.

 \begin{figure}[t]
     \centering
     \vspace{-0mm}
     \includegraphics[width=0.50\textwidth]{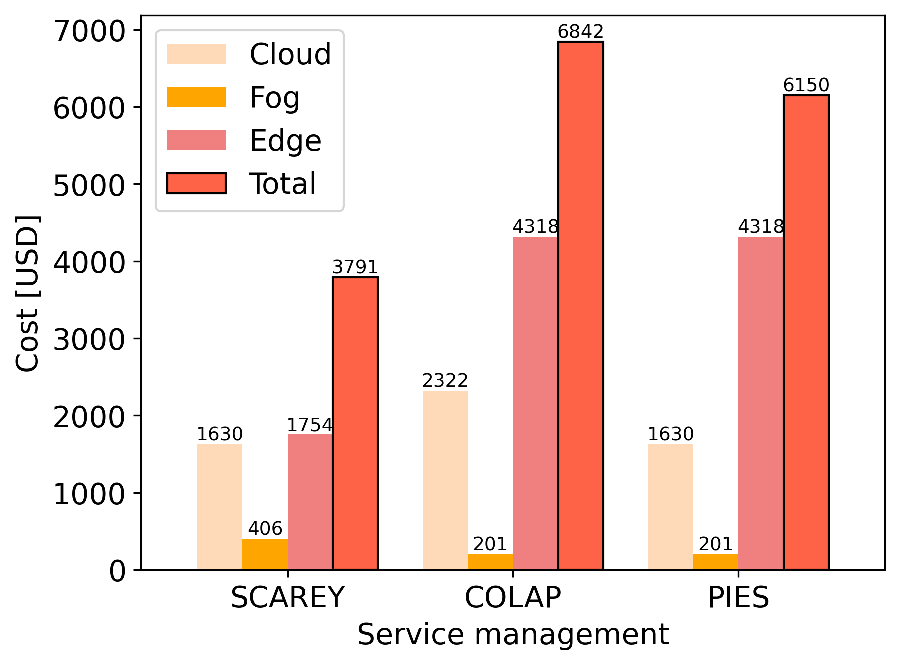} %
     \caption{Service costs}%
     \label{fig:costs}%
         \vspace{-4mm}
\end{figure}

\subsection{$CO_2$ Emissions}\label{sub:emissions}
In Figure \ref{fig:co2with}, we evaluate the $CO_2$ emissions for the service management methods defined in the annual user demand scenario. The total $CO_2$ emissions ($E_{total}$) for SCAREY are \SI{95}{\kilogram}, with contributions of \SI{42}{\kilogram} from the Cloud, \SI{9}{\kilogram} from the Fog, and \SI{43}{\kilogram} from the Edge nodes. In contrast, COLAP generates higher emissions of \SI{180}{\kilogram} annually, comprising \SI{65}{\kilogram} from the Cloud, \SI{9}{\kilogram} from the Fog, and \SI{105}{\kilogram} from the Edge. PIES has total emissions of \SI{157}{\kilogram}, with \SI{42}{\kilogram} from the Cloud, \SI{9}{\kilogram} from the Fog, and \SI{105}{\kilogram} from the Edge nodes.

The reduced $CO_2$ emissions SCAREY achieves are primarily due to its ability to exploit a reduced runtime by dynamically managing resources. SCAREY identifies idle Edge instances without active users and suspends them, decreasing the environmental footprint by limiting the operational hours of the infrastructure. Therefore, SCAREY minimizes the emissions by 47\% through effective runtime optimization.

The differences in Cloud emissions between COLAP and PIES are also notable. COLAP is more CPU-intensive and requires additional resources, leading to using an EC2 \texttt{t4g.2xlarge} instance, contributing to the higher Cloud emissions of \SI{65}{\kilogram}. In contrast, PIES, which demands fewer resources, uses a \texttt{t2.xlarge} instance, resulting in lower Cloud emissions of \SI{42}{\kilogram}. This distinction in resource requirements further underscores the environmental benefits of using more efficient orchestration methods. Figure \ref{fig:co2without} displays the same trends but without emissions due to provided by the manufacturing \cite{BenjaminDavy2021BuildingDataset}.
\begin{figure}%
    \centering
    \subfloat[\centering With manufacturing emmisions]{{\includegraphics[width=0.23\textwidth]{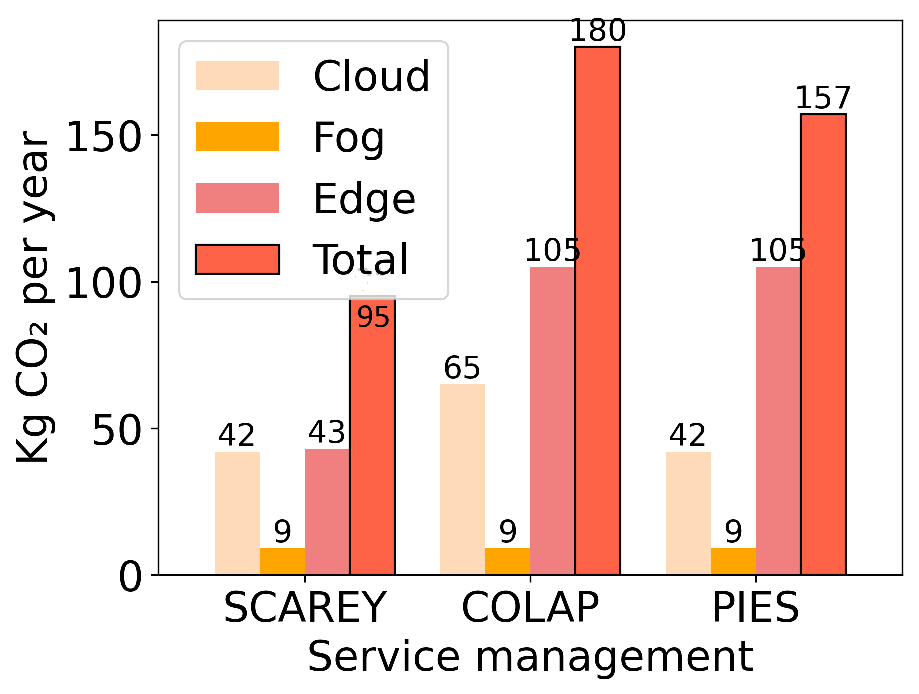} }\label{fig:co2with}}%
    \,
    \subfloat[\centering Without manufacturing emissions]{{\includegraphics[width=0.23\textwidth]{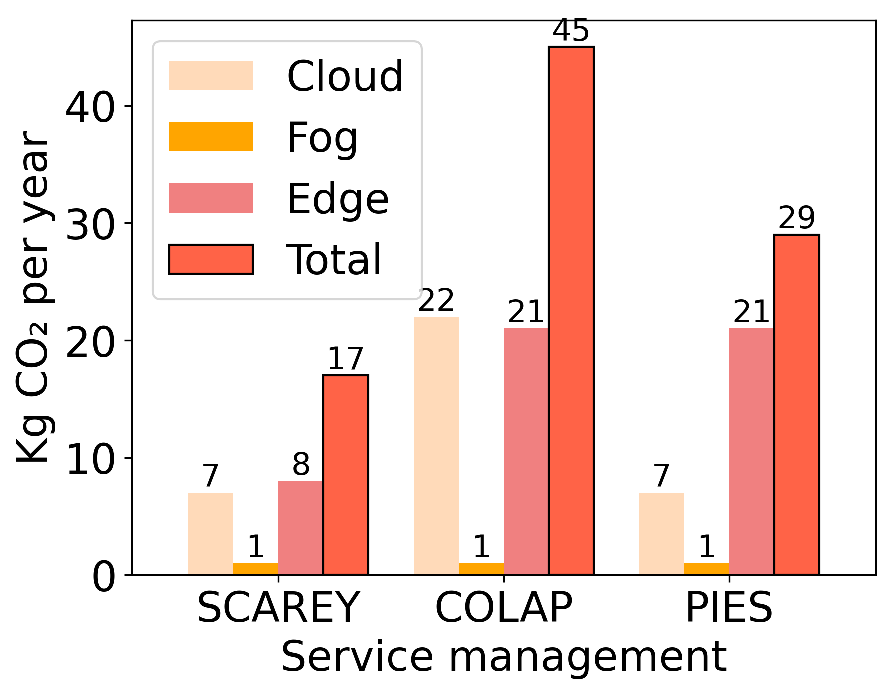} }\label{fig:co2without}}%
    \caption{$CO_2$ emissions}%
    \label{fig:co2}%
    \vspace{-4mm}
\end{figure}



\subsection{Power Consumption}\label{sub:power}
Figure \ref{fig:power} shows the power consumption in kWh per year across service management tools, highlighting SCAREY’s efficiency in reducing energy use. The evaluation is based on scenario annual user demand described in Paragraph \ref{par:scenAnnualdemand}. SCAREY records a total power consumption ($P_{total}$) of \SI{205}{\kilo\watt\hour}, with \SI{59}{\kilo\watt\hour} from the Cloud, \SI{28}{\kilo\watt\hour} from the Fog, and \SI{117}{\kilo\watt\hour} from the Edge. In comparison, COLAP’s total consumption is higher at \SI{486}{\kilo\watt\hour}, consisting of \SI{169}{\kilo\watt\hour} from the Cloud, \SI{28}{\kilo\watt\hour} from the Fog, and \SI{288}{\kilo\watt\hour} from the Edge nodes. Similarly, PIES has an annual power consumption of \SI{376}{\kilo\watt\hour}, with \SI{59}{\kilo\watt\hour} from the Cloud, \SI{28}{\kilo\watt\hour} from the Fog, and \SI{288}{\kilo\watt\hour} from the Edge.

The reduced power consumption achieved by SCAREY is primarily due to its dynamic resource management, allowing it to suspend idle Edge instances when no active users are present, thus limiting operational hours and reducing energy use by 57\%. COLAP requires more computational resources in the Cloud, which leads it to use an EC2 \texttt{t4g.2xlarge} instance, resulting in a higher power consumption of \SI{169}{\kilo\watt\hour}. Conversely, PIES, with less intensive resource requirements, operates on a \texttt{t2.xlarge} instance, resulting in a lower power consumption in the Cloud of \SI{59}{\kilo\watt\hour}. These distinctions in power demands further underscore the energy savings achievable with more efficient orchestration methods like SCAREY.
 \begin{figure}[t]
     \centering
     \vspace{-0mm}
     \includegraphics[width=0.50\textwidth]{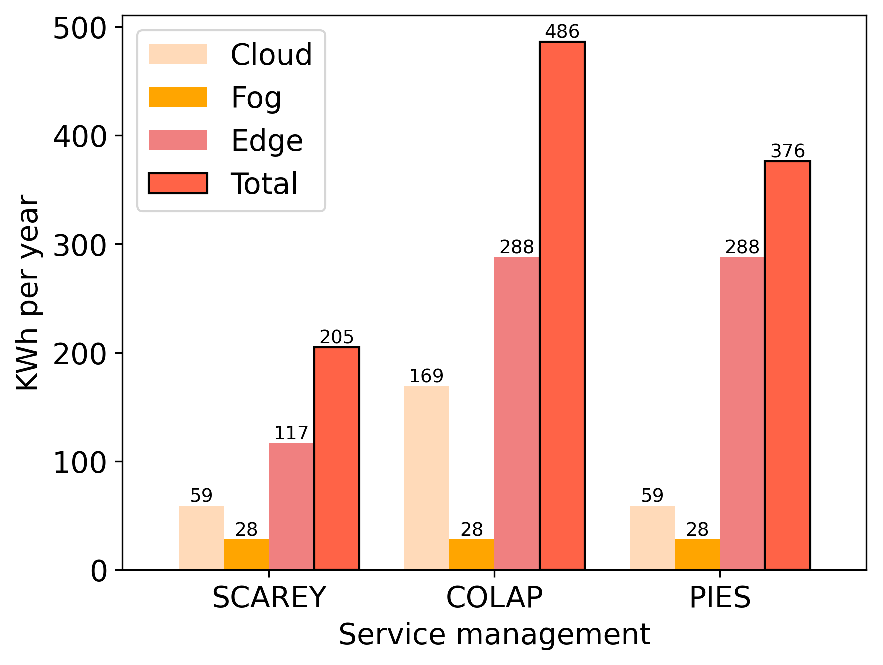} %
     \caption{Power consumption}%
     \label{fig:power}%
         \vspace{-4mm}
 \end{figure}
 
\vspace{-2mm}
\section{Conclusion}\label{sec:Conclusion}
The dynamic and multifaceted nature of scheduling services in the computing continuum requires innovative and comprehensive solutions. In this paper, we introduced SCAREY, a user location-aided service lifecycle management approach based on state machine modelling. SCAREY addresses key challenges in service discovery, provisioning, placement, and monitoring by offering a unified framework that dynamically adapts to fluctuating user demand and resource availability.

The proposed method demonstrates significant contributions to service lifecycle management through the state machine-based approach, enabling demand-driven transitions, scalable deployment, and resource optimization. By leveraging network measurements for service placement and implementing selective resource acquisition and release strategies, SCAREY ensures efficient service delivery, reduced latency, and improved sustainability by minimizing resource wastage and carbon emissions. 

To validate the performance of SCAREY, we conducted real-world evaluations across multiple scenarios and compared its performance against two related frameworks. The results demonstrated that SCAREY significantly enhances service discovery and acquisition time, achieving improvements of over 73\%. Additionally, it reduces operating costs by 45\% and power consumption by more than 57\% compared to the alternative approaches.

Future research will focus on extending this framework to incorporate additional QoS metrics, address security concerns, and explore its applicability in emerging domains, such as autonomous systems and 6G networks, to further enhance its versatility and impact.

\section*{Acknowledgements}
This work received funding from the Austrian Research Promotion Agency (FFG), project 888098 (K{\"a}rntner Fog) and project 909989 (AIM AT Stiftungsprofessur f\"ur Edge AI).

\bibliographystyle{IEEEtran}
\bibliography{external}

\end{document}